\documentclass[12pt]{article}

\usepackage{amssymb, amsmath, enumerate, theorem, epsfig}

\begin{document}

\mathchardef\bsurd="1371
\mathchardef\bsolid="132E

\def\dd#1#2{{{\rm d} #1\over {\rm d} #2}}
\def\ddn#1#2#3{{{\rm d}^#3 #1\over {\rm d} #2^#3}}
\def\bb#1{{\bf #1}}
\def\pd#1#2{{\partial #1\over\partial #2}}
\def\bk{{\bf k}}
\def\bv{{\bf v}}
\def\bw{{\bf w}}
\def\bx{{\bf x}}
\def\bz{{\bf z}}

\def\pmb#1{\setbox0=\hbox{#1}%
\kern-.025em\copy0\kern-\wd0
\kern.05em\copy0\kern-\wd0
\kern-.025em\raise.0433em\box0}

\def\bze{{\pmb{$0$}}}
\def\bu{{\pmb{$1$}}}
\def\sm{{\pmb{$\cdot$}}}
\def\vm{{\pmb{$\times$}}}
\def\grad{{\pmb{$\nabla$}}}
\def\div{{\grad\sm}}
\def\curl{{\grad\vm}}
\def\bom{{\pmb{$\omega$}}}
\font\smallrm=cmr8 scaled \magstep 0

\title{\bf Motion in a Bose Condensate: IX. Crow instability of antiparallel 
vortex pairs}
\author{Natalia G. Berloff and  Paul H. Roberts\\
Department of Mathematics\\
University of California, Los Angeles, CA, 90095-1555\\ 
{\it nberloff@math.ucla.edu, roberts@math.ucla.edu}}
\date {July 26, 2001}
\maketitle
\begin {abstract}
The Gross-Pitaevskii (GP) equation admits a two-dimensional solitary wave solution
representing two mutually self-propelled, anti-parallel straight line vortices. 
The complete sequence of such solitary wave solutions has been computed by 
Jones and Roberts (J. Phys.\ A, {\bf 15}, 2599, 1982). These solutions 
are unstable with respect to three-dimensional perturbations (the Crow instability).
The most unstable mode has a wavelength along the direction of the vortices
of the same order as their separation. The growth rate associated with this
mode is evaluated here, and it is found to increase very rapidly with decreasing
separation. It is shown, through numerical integrations of the GP equation that,  
as the perturbations
grow to finite amplitude, the lines reconnect to produce a sequence of
almost circular vortex rings.
\end{abstract} 

\section{Introduction}

The experimental realization of Bose condensates in weakly interacting gases
motivated the recent explosion in theoretical studies of its properties
using the the Gross-Pitaevskii (GP) model which represents the so-called mean-field
limit of quantum field theories. The same equation has been the subject of
extensive studies also in the framework of superfluid helium at very low
temperature. In this case the GP model is assumed to be linked to the
condensate fraction of the superfluid. This is the ninth in a series of papers 
devoted to modeling flows in a
Bose Condensate. Reference will  be made to the fourth and fifth  papers in the
sequence: Jones and Roberts (1982) and Jones, Putterman, and Roberts (1986).

Superfluid turbulence has been the focus of many experimental and
theoretical studies (see Donnelly, 1991). Superfluid turbulence manifests 
itself as a tangle of quantized vortex lines. The dynamics of the tangle 
depend crucially on the interactions of the vortex filaments. 
These have been studied by using the classical model of vortices
in an incompressible Euler fluid. This omits, however, two
mechanisms that are very relevant to the superfluid tangle. 

First, as Vinen (2000)  argues,   emission of sound by a 
vortex tangle is very significant in superfluid turbulence. This process
is completely removed by the main assumption of classical
vortex theory: $\div\bv=0$, where $\bv$ is the superfluid velocity. 
The dynamics of vortex filaments 
in a compressible fluid is not as well understood as that for
the incompressible case. The scattering of sound by compressible
Euler fluids has however been the subject of several
recent investigations; see, for example, Ford and Smith (1999).

Second, the processes of severance and coalescence of vortex lines
are centrally important for the study of superfluid turbulence,
but these are expressly forbidden by the Kelvin-Helmholtz 
theorem, according to which vortex lines are frozen to an 
Euler fluid and cannot change their topology. In an Euler  fluid, the 
processes have been successfully simulated numerically by
restoring viscosity. This step is disallowed
in a superfluid, and the only way to defeat the theorem
is through {\it ad hoc} procedures. For example, it was supposed by
Schwarz (1988) that, whenever one vortex filament comes within a distance 
$\Delta$ of another filament, reconnection will 
always occur, and that otherwise reconnection will not happen. 
A precise way of determining $\Delta$ is not known, but its value 
can clearly greatly affect the reconnection rate in a vortex tangle.
Moreover, the angle at which the vortex filaments approach one
another is undoubtedly an important factor in determining
whether they reconnect or not; a clear set of reconnection rules
is lacking.  

The advantage of GP theory in comparison with the classical approach is 
that it gives superfluid vortex lines their own unique core structure. 
At the same time, it provides a mechanism for the severance and 
coalescence of vortex lines, and includes sound propagation, 
so that the acoustic emission from a vortex tangle can be evaluated.
Koplik and Levine (1993, 1996)  used numerical simulations 
of the GP model to study the reconnection of, and the interaction 
between, straight-line vortices and vortex rings. In 
particular, they witnessed the annihilation of  vortex rings of similar radii. 
Recently  Leadbeater {\it et al.} (2001) elucidated the loss of energy 
to sound emission during vortex ring collisions. Their calculations suggested 
that the sound emitted during reconnections is a significant decay mechanism 
for superfluid turbulence.

In this paper, we first study the linear stability of a vortex pair.
This is a two-dimensional (2D) structure that can, in GP theory, be
represented by a wavefunction $\psi_0(x,y,t)$ that is independent of the
coordinate $z$ and which has two zeros at $y=\pm{\tfrac{1}{2}}h$, representing
vortices separated by a distance $h$. The phase of $\psi_0$ increases by
$2\pi$ round one zero and decreases by $2\pi$ round the other corresponding,
in the hydrodynamic interpretation of $\psi_0$, to a pair of antiparallel
vortices (sometimes called ``point vortices") that move uniformly with
speed $U$ in the $x-$direction as a solitary wave, i.e., $\psi_0=\psi_0(x-Ut,y)$,
where $U$ is obtained by solving the GP equation in 2D:
\begin{equation}
2 {\rm i} U {\partial \psi_0 \over \partial x} 
= {\partial^2  \psi_0 \over \partial x^2} 
+{\partial^2  \psi_0 \over \partial y^2}  + (1 - |\psi_0|^2)\psi_0.
\label{Ugp}
\end{equation}
(Here, and frequently in what follows, we use dimensionless variables
 such that the unit length corresponds to the healing
length $a$, the speed of sound is $c=1/\sqrt{2}$,  and the density at infinity 
is $\rho_\infty=1$. Later we shall write $\psi_0=u_0+{\rm i}v_0$, where $u_0$ 
and $v_0$ are real.)
Solutions of this form were first reported by Jones and Roberts (1982) 
who determined the entire of sequence of such solutions and their
associated energy per unit length $\cal E$ and momentum per unit length ${\cal P}$,
both of which decrease to zero as $h\to0$:
\begin{eqnarray}
{\cal E}&=&{\tfrac{1}{2}}\int|\nabla \psi_0|^2\, dV + {\frac{1}{4}}\int (1 - |\psi_0|^2)^2\, dV \label{eJR}\\
{\cal P}&=&{\frac{1}{2{\rm i}}}\int [(\psi_0^*-1)\nabla\psi_0-(\psi_0-1)\nabla\psi_0^*]\, dV.
\label{pJR}
\end{eqnarray}
Multiplying (\ref{Ugp}) by $x{\partial \psi_0^*/\partial x}$ and integrating 
by parts, Jones {\it et al} (1986)  showed that
\begin{equation}
{\cal E}={\tfrac{1}{2}}\int\biggl|{\pd {\psi_0} x}\biggr|^2\, dV.
\label{eJR2}
\end{equation}
They located a critical value $h_c \approx 1.7$ of $h$, at which the sequence 
lost or gained vorticity. For $h<h_c$  the sequence has no vorticity, although 
solitary disturbances exist as
finite amplitude sound waves in which the two minima of $|\psi_0|$ are no longer 
zero.
As $h\to 0$, $U$ approaches the speed of sound $c$ and the acoustic solutions merge 
with the phonon branch of the dispersion curve. 

Jones and Roberts did not examine the stability of their 2D solitary waves.
It is known that the vortex pair in an incompressible Euler fluid
is prone to the so-called ``Crow instability" (Crow, 1970). Kuznetsov and 
Rasmussen (1995) proved that  in the long-wavelength limit, where $k$ is small 
compared with $h$, both the vortex pair  and solitary acoustic  solutions are 
unstable, but they   determined neither the boundaries of instability nor the  
wavelength at which the growth rate is maximal.  In this paper we first solve 
the linear stability problem for all $h$
with the particular aim of finding the growth rate of the Crow instabilities
as a function of the separation $h$. We study the 
subsequent evolution of the instabilities to finite amplitude by integrating
the GP equations in 3D. This parallels the corresponding analysis by
Moore (1972) for a classical fluid, but differs in that
healing becomes important as the instability brings one vortex core
close to the other. Unlike the classical case, reconnection can, and does,
occur so that the final result is a sequence of almost circular vortex rings.
\section{Linear stability of the vortex pair}

We return to the GP equation  in the reference frame moving
with the vortex pair:
\begin{equation}
-2 {\rm i} \pd {\psi} t + 2 {\rm i} U {\partial \psi \over \partial x} 
= \nabla^2\psi  + (1 - |\psi|^2)\psi.
\label{tUgp}
\end{equation}
We  seek solutions of (\ref{tUgp}) in the form 
$\psi(x,y,z,t) = \psi_0(x,y) + \widehat\psi(x,y,z,t)$ 
where $\widehat\psi$ is infinitesimal. The resulting linearized GP
equation determines the stability of the vortex pair. We separate 
$\widehat \psi$ into real and imaginary parts, $\widehat u$ and $\widehat v$, 
and focus on separable solutions of the form
\begin{equation}
\widehat u = u(x,y)\exp[\sigma t - ikz]+u^*(x,y)\exp[\sigma t + ikz],
\label{uhat}
\end{equation}
and similarly for $\widehat v$, where $*$ stands for complex conjugation; 
the functions $u$ and $v$ are governed by
\begin{eqnarray}
 \nabla_{xy}^2 u +2 U {\partial v \over  \partial x} 
+  \Bigl(1 - 3 u_0^2 -v_0^2 - k^2\Bigr)u -2u_0v_0v &=&  \quad2\sigma v,\label{stability1}\\
 \nabla_{xy}^2 v -2 U {\partial u \over  \partial x} 
+  \Bigl(1 -  u_0^2 - 3v_0^2 - k^2\Bigr)v -2u_0v_0u &=& - 2\sigma u,
\label{stability2}
\end{eqnarray}
and, since the perturbation must vanish at great distances from the vortex 
pair, we have
\begin{equation}
u\rightarrow 0, \quad v\rightarrow 0, \quad {\rm for} \quad s\equiv\sqrt{x^2+y^2}\rightarrow \infty.
\label{atinf}
\end{equation}

The linear stability problem posed by (\ref{stability1})-(\ref{atinf}) 
has features in common with the corresponding classical stability problem analyzed 
by Crow (1970). In particular, it follows from  (\ref{stability1})-(\ref{atinf}) 
that $\sigma^2$ is real. This may be demonstrated by introducing adjoint variables 
$\bar u$ and $\bar v$ that obey (\ref{atinf}) and share the same eigenvalue 
spectrum. We multiply (\ref{stability1}) by $\bar u$, (\ref{stability2}) 
by $\bar v$, add corresponding sides, integrate over the interior of the 
cylinder $s=S$, apply the divergence theorem, discarding the resulting surface 
integrals for $S\rightarrow \infty$ by an appeal to (\ref{atinf}). We then find 
that $\bar u$ and $\bar v$ must obey (\ref{stability1})-(\ref{atinf}), but with 
$\sigma$ replaced by $-\sigma$. In short, if $\sigma$ is an eigenvalue of 
(\ref{stability1})-(\ref{atinf}), so is $-\sigma$. Since all coefficients 
in (\ref{stability1}) and (\ref{stability2}) are real, $\sigma$ and $\sigma^*$ 
are both eigenvalues. Thus in all cases $\sigma^2$ is real.

The eigenvalues of (\ref{stability1})-(\ref{atinf}) belong to two distinct 
types  of instability, termed the symmetric and the antisymmetric modes:
\begin{eqnarray}
{\rm symmetric:} \quad u(-x,y)&=&-u(x,y),\quad v(-x,y)=v(x,y)\nonumber\\
{\rm antisymmetric:} \quad u(-x,y)&=&u(x,y),\qquad v(-x,y)=-v(x,y).\nonumber
\end{eqnarray}
Kuznetsov and Rasmussen (1995) demonstrated that all long wavelength 
antisymmetric modes are stable and all long wavelength symmetric modes 
are unstable. In fact they showed that the dispersion relation for the 
antisymmetric perturbation is 
\begin{equation}
\sigma^2 = (kU)^2\biggl(1 - {\frac{{\cal E}}{{\cal P}U}}\biggr) < 0, \quad k\rightarrow 0,
\label{asym}
\end{equation}
and the growth rate of symmetric perturbation is given by
\begin{equation}
\sigma^2 =  - {\frac{{\cal E}}{\partial{\cal P}/\partial U}} k^2 > 0, \quad k\rightarrow 0,
\label{sym}
\end{equation}
where ${\cal E}$ and ${\cal P}$ are the energy and momentum per unit length of 
the vortex pair; see (\ref{eJR}) and (\ref{pJR}). These have been evaluated by 
Jones and Roberts (1982) and by Jones, Putterman and Roberts (1986) for the 
entire vortex sequence, from the KP1-soliton for ${\cal P}\rightarrow 0$ to 
a widely separated pair of  vortices for ${\cal P}\rightarrow \infty$. In the 
latter  case,  it was found that, in dimensional units, 
\begin{equation}
{\cal E}\sim {\frac{\rho_\infty \kappa^2}{2\pi}} \biggl[\ln {{\frac{h}{a}}}+\alpha\biggr],
\label{en}
\end{equation}
\begin{equation}
{\cal P}\sim\rho\kappa h, \qquad U  \sim {\frac{\kappa}{2\pi h}},
\label{pu}
\end{equation}
where $a$ is the healing length and
$\alpha$ is the vortex core parameter determined numerically by Pitaevskii (1961) 
as $\alpha \approx 0.38$. To compare (\ref{sym}) with the result obtained by Crow 
(1970) we rewrite (\ref{en})  using the cut-off method (Saffman, 1992). According 
to this method we estimate the vortex cut-off parameter $\delta$ by comparing the 
velocity of a ring of radius $R$ given by the cut-off formula
\begin{equation}
U={\frac{\kappa}{16\pi R}}\int_{a\delta /R}^{2\pi -a\delta /R} {\frac{1}{ \sin{\tfrac{1}{2}}\theta}}\, d\theta = {\frac{\kappa}{4\pi R}} \ln {\frac{4 R}{a\delta }}
\label{U1}
\end{equation}
with the analytical result (Grant and Roberts, 1971)
\begin{equation}
U={\frac{\kappa}{4\pi R}}\biggl(\ln {\frac{8R}{a}}-1+\alpha\biggr).
\label{U2}
\end{equation}
This comparison gives us 
\begin{equation}
\ln 2 \delta = 1-\alpha.
\label{deltaGP}
\end{equation}
Using (\ref{deltaGP}) as a definition of the cut-off parameter for the GP model 
we can write (\ref{en}) as
\begin{equation}
{\cal E}\sim {\frac{\rho_\infty \kappa^2}{2\pi}} \biggl[\ln {{\frac{h}{2 a\delta}}}+1\biggr],
\label{newen}
\end{equation}
which together with (\ref{pu}) and (\ref{sym}) implies that, for symmetric modes,
\begin{equation}
\sigma^2\sim\biggl({ \frac{\kappa}{2 \pi h}}\biggr)^2k^2\biggl[\ln{\frac{h}{2 a\delta}}+1\biggr], \quad k\rightarrow 0.
\label{sig2}
\end{equation}
This establishes they are unstable for all  sufficiently large wavelengths.

It is possible here to compare (\ref{sig2}) with the classical theory of 
Crow (1970), in which $\delta$ is the cut-off employed when vorticity is 
assumed to be confined to filaments. Crow assumed the uniform core vortex 
model, but his derivation is easily adapted to a vortex pair with other  core 
structures. He found that, provided $ka\ll 1$, where $a$ is the core radius,
\begin{eqnarray}
\sigma^2=\biggl({\frac{\kappa}{2 \pi h^2}}\biggr)^2 \biggl[&1& + khK_1(kh)-{\tfrac{1}{2}}(kh)^2\omega(\delta)\biggr] \nonumber \\
\times\biggl[&1& - khK_1(kh)-(kh)^2K_0(kh)+{\tfrac{1}{2}}(kh)^2 \omega( \delta)\biggr],
\label{sss}
\end{eqnarray}
where $K_0$ and $K_1$ are modified Bessel functions, and 
\begin{equation}
\omega(\delta)=-2\int_{ak\delta}^\infty( \cos u + u\sin u -1){\frac{du}{u^3}} \sim \ln(a k \delta)  + \gamma -{\tfrac{1}{2}}+{\cal O}(\delta)^2.
\end{equation}
Here $\gamma\approx0.577216\cdots$ is Euler's constant. When we approximate 
(\ref{sss}) for $kh\ll 1$, we obtain (\ref{sig2}).

Now we will address the question of whether   the expression (\ref{sym}) has 
a more general meaning and is valid for the classical core models, so that we 
can adopt  (\ref{sig2}) as the general expression  for the  growth  rate of 
large wavelength perturbations. For the uniform core model we relate (\ref{U1}) 
to the analytical expression for the velocity of a  vortex ring of radius $R \gg a$:
\begin{equation}
U={\frac{\kappa}{4\pi R}}\biggl(\ln {\frac{8R}{a}}-{\frac{1}{4}}\biggr).
\label{U3}
\end{equation}
This comparison defines the cut-off parameter as $ 2\delta = {\rm e}^{1/4}$. 
The energy of two antiparallel uniform core vortices is 
\begin{equation}
{\cal E}\sim {\frac{\rho_\infty \kappa^2}{2\pi}} \biggl[\ln {{\frac{h}{ a}}}+{\frac{1}{4}}\biggr],
\label{ucve}
\end{equation}
which, when written using the cut-off parameter, becomes
\begin{equation}
{\cal E}\sim {\frac{\rho_\infty \kappa^2}{2\pi}} \biggl[\ln {{\frac{h}{2 a\delta}}}+{\frac{1}{2}}\biggr].
\label{ucve2}
\end{equation}
The momentum and velocity of the vortex pair are given by (\ref{pu}). 
When the right-hand side of the expression (\ref{sym}) is evaluated using 
expressions (\ref{pu}) and (\ref{ucve2}) the result becomes 
\begin{equation}
 - {\frac{{\cal E}}{\partial{\cal P}/\partial U}} k^2 = \biggl({ \frac{\kappa}{2 \pi h}}\biggr)^2k^2\biggl[\ln{\frac{h}{2 a\delta}}+{\frac{1}{2}}\biggr],
\label{ucres}
\end{equation}
which differs from (\ref{sig2}) by  ${\tfrac{1}{2}}$. The nature of this 
difference together with a brief description of the cut-off method are given 
in the Appendix.

The linear stability problem (\ref{stability1})-(\ref{atinf}) was solved 
numerically for various $h$. The region of instability and the maximum 
growth rate were determined in the $kh$-plane; see Figure 1.
The stability boundary for the classical hollow core  vortices is depicted 
in Figure 1 as well.
\begin{figure}

\caption{\small Stability boundary (dots and solid lines - for superfluid 
solitary waves, bold  solid line -- for classical hollow core vortices) 
and maximum growth rate (dashed line) for (\ref{stability1})-(\ref{atinf})}
\bigskip
\bigskip
\centering
\psfig{figure=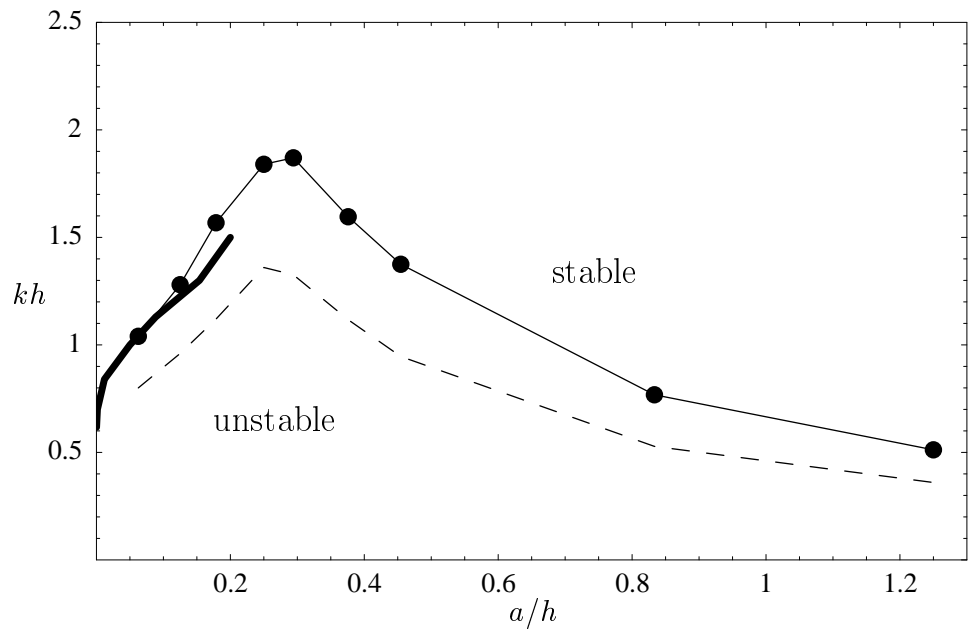,height=3.4in}
\end{figure}
\bigskip  

\vfill\eject
\section{Nonlinear evolution of the instability}

As the unstable perturbation grows in amplitude, it can no longer
be described by linear equations such as (\ref{stability1})-(\ref{atinf}).
To determine its subsequent evolution it is necessary to undertake
direct numerical integrations of the GP equation. One can then
understand the evolution in the following way. As the instability grows, 
it brings some segments of one line into closer proximity with
corresponding segments of the other line, until the minimum distance
between the lines reaches a critical value, approximately equal to the
critical value ($h=h_c$) at which vorticity is lost on the solitary
wave sequence; see \S 2.
At this moment, the Kelvin-Helmholtz theorem is inapplicable and 
reconnection occurs; curves of zero $\psi$ on one line join with 
the curves of zero $\psi$ on the other to form closed elongated
vortex rings that later relax to become approximately circular.  Before doing so, 
each ring oscillates in its fundamental mode, being alternatively prolate and 
oblate; the amplitude of this oscillation diminishes as it radiates acoustic waves.

This scenario is supported by direct numerical simulations,
performed with the same numerical method as in our previous work 
(Berloff and Roberts, 2000). 
In these computations we follow 
the evolution of a vortex pair moving in the $x-$direction
in a computational box of dimensions $D_x=60$, $D_y=60$, $D_z=120$. 
The $xy-$faces of the box are open, to allow sound waves
to escape; this is achieved numerically by applying the Raymond-Kuo
technique (Raymond and Kuo, 1984). The faces $z=0$ and $z=D_z$ are reflective.
To introduce an initial perturbation that does not favor any particular 
wavelength we start with the initial condition 
\begin{equation}
\psi(x,y,z,t=0)=\psi_0(x,y-3)*\psi_0(x,y+3),
\label{ic}
\end{equation}
 where 
\begin{equation}
\psi_0(x,y)=[1-\exp(-0.7r^{1.15})]\exp({\rm i}\theta)
\label{ic2}
\end{equation}
 is an  approximation for the rectilinear vortex and $r$ and $\theta$ are 
polar coordinates, such that $x=r\cos \theta$ and $y=r\sin\theta$. 
The
wavelength of the instability for $h=6$ was about $30$, corresponding 
to $k\approx 0.2$, in good agreement with the result of the linear 
stability analysis, which gave $k\approx 0.19$. This wavelength determines 
where the vortex filaments approach each other and reconnect as 
vortex rings; see Figure 2.  

The reconnections are accompanied by the 
emission of sound waves and rarefaction pulses, resulting in line loss 
that is approximately  one fourth of the total vortex line length, so confirming 
that acoustic losses are significant, and should to be taken into account 
when modeling superfluid turbulence.

\begin{figure}

\caption{ \small The isosurface $\rho/\rho_\infty = 0.2$ for two anti-parallel 
vortices initially distance $h=6$ apart that propel each other away from the 
viewer. An instability develops
along the axes of these vortex lines, and the lines reconnect to form 
circular vortex rings. }
\bigskip
\bigskip
\centering
\psfig{figure=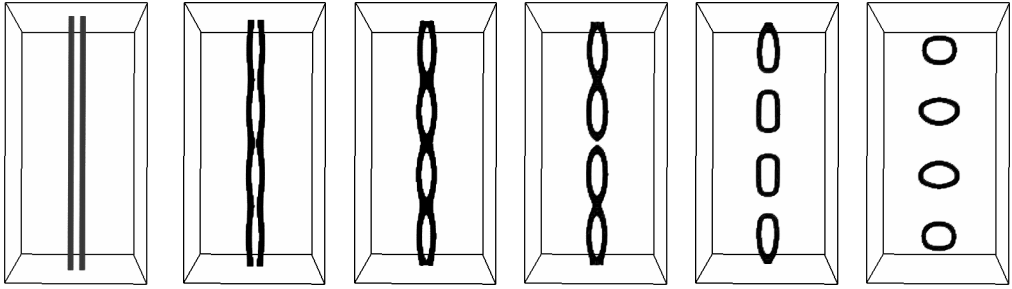,height=1.6in}
\end{figure}
\bigskip

The results of our computations for other $h$ are summarized  in Table 1 which 
gives energy, momentum, and wavelength $\ell$ of the perturbation for the initial 
field and energy, momentum, and radius of the resulting vortex ring and the amount 
of  vortex line lost as a percentage of the initial vortex line length.
To reduce the time taken by an initial perturbation to grow, we took the initial 
$\psi$ to be $\psi_0(x,y_++h/2)*\psi_0(x,y_-h/2)$ where $\psi_0$ is given by 
(\ref{ic2}),   $y_\pm= y\pm 0.1 \cos k z$ and $\psi_0$  and $k$ is the wavenumber 
for which the growth rate is a maximum according to the linear theory of \S 2.
Notice that if $h$ is  small (though larger than the critical distance for 2D 
vortex pair)   the vortices  annihilate each other and the energy is carried 
away as sound waves with the intermediate formation of 3D rarefaction pulses. 

\bigskip

\noindent {\bf Table 1.}

\bigskip
\begin{tabular}{|c|c|c|c|c|c|c|c|}
\hline
$h$ &${\ell}$ & ${\cal E}_{{\rm init}}$ & ${\cal P}_{{\rm init}}$ &$ {\cal E}_{{\rm ring}}$ & ${\cal P}_{{\rm ring}}$ & $ R_{\rm ring}$ & \% of line lost \\ \hline
 8    & 53   & 819 & 2664 & 663 & 2612 & 11.5 & 32    \\
 6    & 33   & 450 & 1244 & 393 & 1204 & 7.8  & 26  \\
 3.6  & 16   & 167 & 362  & 125 & 228  & 3.3  & 35  \\
 2.2  & 14.6 & 107 & 201  & --  & --   & --   & 100  \\\hline
\end{tabular} 
\bigskip

\section{Conclusions}
We studied the Crow instability for two antiparallel vortex lines obeying the 
GP equation. 
Linear stability analysis was used to determine the maximum growth rate of the 
instability and the region of  instability.
Through numerical simulations of the GP equation, we showed   that as 
perturbations grow to finite amplitude the lines reconnect to produce 
a sequence of almost circular vortex rings. We evaluated the resulting 
line loss.

\section*{Acknowledgments}
We are grateful to Dr. Sergey Nazarenko for useful discussions.
This work is supported by the NSF grants DMS-9803480 and DMS-0104288.
\section*{Appendix. Classical Crow instability.}
Crow (1970) used the cut-off method to determine the growth rate, $\sigma$, 
of the instability in an incompressible fluid, for all $kh$ and for $ka\ll 1$, 
where $a$ is the radius of the vortex core. This approximate method is based 
on the Biot-Savart law determining the fluid velocity, ${\bf v({\bf x})}$, 
in an incompressible fluid from an assigned vorticity \bom$({\bf x})$:
\begin{equation}
\bb v(\bb x) = {\frac{1}{4 \pi}}\int{\frac{\bom(\bb x')\times (\bb x- \bb x')}{|\bb x- \bb x'|^3}}\, d\bb x'.
\label{A1}
\end{equation}
It is supposed that the vorticity is confined to filaments of infinitesimal 
cross-section so that (\ref{A1}) reduces to a line integral
\begin{equation}
\bb v(\bb x) = {\frac{\kappa}{4 \pi}}\int{\frac{d\bb s'\times (\bb x- \bb x( s'))}{|\bb x- \bb x(s')|^3}},
\label{A2}
\end{equation}
where $s$ is arc length on a filament and $\kappa$ is the vorticity contained 
within it.

To determine the velocity, $\bb U(\bb x)$, of the filament it is necessary 
to evaluate (\ref{A2}) for each point on the filament $\bb x(s)$, but the 
resulting integral (\ref{A2}) 
diverges. In the cut-off method the offending segment $|\bb s-\bb s'|<a\delta$ 
of the integral is arbitrarily removed, where $\delta$ is the cut-off parameter 
the value of which depends on the core structure. This step is denoted 
by $[\delta]$: 
\begin{equation}
\bb U(s) = {\frac{\kappa}{4 \pi}}\int_{[\delta]}{\frac{d\bb s'\times (\bb x(s)- \bb x( s'))}{|\bb x(s)- \bb x(s')|^3}}.
\label{A3}
\end{equation}

We give two examples. First for our vortex pair, no divergence arises because 
$d\bb s'$ and $\bb x(s)-\bb x(s')$ are parallel when $s'$ and $s$ are on the 
same line. Thus
\begin{equation}
U = {\frac{\kappa h}{4 \pi}}\int_{-\infty}^{\infty}{\frac{d s}{(h^2+s^2)^{3/2}}}={\frac{\kappa}{2 \pi h}}.
\label{A4}
\end{equation}
In the second case, a vortex ring of radius $R (\gg a)$, (\ref{A3}) gives (\ref{U1}).

An alternative way of defining a cut-off is through the expression for the 
energy of a vortex line assembly. This is most conveniently expressed as 
in \S 153 of Lamb (1945) as
\begin{equation}
E={\frac{\rho}{8 \pi}}\int\int{\frac{\bb\omega(\bb x)\sm \bb\omega(\bb x') }{|\bb x- \bb x'|}}\,d\bb x d\bb x',
\label{A6}
\end{equation}
which, when the vorticity is concentrated into filaments is, 
\begin{equation}
E={\frac{\rho \kappa \kappa'}{8 \pi}}\int_{[\bar\delta]}\int_{[\bar\delta]}{\frac{d\bb s\cdot d\bb s' }{|\bb x(s)- \bb x(s')|}}\,d\bb x d\bb x',
\label{A7}
\end{equation}
and $[\bar \delta]$ signifies that the segment $|s'-s|<a\bar \delta$ is 
removed. Returning to our two examples, (\ref{A7}) gives
\begin{equation}
{\cal E} = {\frac{\rho \kappa^2}{2 \pi}} \ln {\frac{h}{2 a \bar \delta}},
\label{A8}
\end{equation}
for the vortex pair and 
\begin{equation}
E={\frac{1}{2}}\rho\kappa^2 R \biggl(\ln {\frac{4 R}{a \bar \delta}} - 2\biggr),
\label{A9}
\end{equation}
for the thin-cored ring.

The cut-offs $\delta$ and $\bar\delta$ must be chosen differently. In order 
that the Hamiltonian relation
\begin{equation}
U=\partial E /\partial P
\label{A10}
\end{equation}
is obeyed by the ring, where $P=\rho \kappa \pi R^2$ is its momentum (impulse), 
it is necessary to hold the volume $2\pi^2Ra^2$ of the ring constant in the 
differentiation (\ref{A10}); see Roberts and Donnelly (1970) and Roberts (1972). 
This requirement, which implies that 
\begin{equation}
\ln(\delta /\bar\delta) = {\tfrac{1}{2}},
\label{A11}
\end{equation}
is relevant even for the hollow core vortex, since any change in volume would 
imply that $pdV$ work is done on the system at infinity, with a concomitant 
change in $E$ that would cause (\ref{A10}) to fail. Relations (\ref{A8}), 
(\ref{A9}), and (\ref{A11}) show that 
\begin{eqnarray}
{\cal E} &=& {\frac{\rho \kappa^2}{2 \pi}} \biggl(\ln {\frac{h}{2 a \bar \delta}}+{\frac{1}{2}}\biggr), \label{A12}\\
E &=& {\frac{1}{2}}\rho\kappa^2 R \biggl(\ln {\frac{4 R}{a \bar \delta}} - {\frac{3}{2}}\biggr).
\label{A13}
\end{eqnarray}
Result (\ref{A13}) agrees with known facts for the uniform core 
$(\ln 2\delta = {\tfrac{1}{4}})$ and hollow core $(\ln 2\delta = {\tfrac{1}{2}})$ 
rings; see Lamb (1945) and Saffman (1992). (The uniform core vortex is one in 
which $\omega/s$ is constant where $\omega \bb 1_\phi$ is the vorticity and 
$\bb 1_\phi$ is the unit vector in the direction of increasing $\phi$; 
see \S 165 of Lamb, 1945.) Similarly for the vortex pair, 
$U=\partial {\cal E}/\partial{\cal P}$, where
\begin{equation}
{\cal P} = \rho \kappa h
\label{A14}
\end{equation}
is the momentum per unit length.

The expressions of Crow for the growth rates of his antisymmetric and 
symmetric modes of instability reduce, in the limit $kh\rightarrow 0$, 
to (\ref{asym}) and (\ref{sym}) above, but {\bf not} with the 
expression (\ref{A12}) for ${\cal E}$. In its place stands
\begin{equation}
\widehat{\cal E} = {\frac{\rho \kappa^2}{2 \pi}} \biggl(\ln {\frac{h}{2 a \bar \delta}}+1\biggr). 
\label{A15}
\end{equation}
It seems to us that this puzzling difference may be connected to the 
different frame of reference used in deriving (\ref{A15}). Crow used 
the co-moving frame, in which (\ref{A15}) translates to the energy 
$\widetilde {\cal E} = \widehat {\cal E} - {\cal P} U$:
\begin{equation}
\widetilde {\cal E} = {\frac{\rho \kappa^2}{2 \pi}} \ln {\frac{h}{2 a \bar \delta}}.
\label{A16}
\end{equation}
For the reason why this does not coincide with (\ref{A12}) we offer 
the following speculation.

In the laboratory frame (the frame in which the fluid is at rest at infinity) 
the streamfunction $\psi(x,y)$ for a hollow core vortex is
\begin{equation}
\psi = {\frac{ \kappa}{2 \pi}}\ln {\frac{r_2}{r_1}},
\label{A17}
\end{equation}
where $r_1$ $[r_2]$ is a distance from $(0, {\tfrac{1}{2}}h)$ 
[from $(0, -{\tfrac{1}{2}}h)$]; since these distances change with time, $\psi$ 
is implicitly a function of $t$. The streaklines, i.e., the curves parallel to 
the instantaneous direction of the velocity $\bb v = \grad \times (\psi\bb 1_z)$ 
are coaxial circles surrounding the cores, the surfaces of which are (for $a \ll h$)
 $r_1=a$ and $r_2=a$. The energy per unit length is (see \S 157 of  Lamb, 1945)
\begin{equation}
{\cal E} = {\tfrac{1}{2}}\rho \int v^2d^3x = {\tfrac{1}{2}}\rho \kappa (\psi_{s1}-\psi_{s2})=\rho\kappa\psi_{s1},
\label{A18}
\end{equation}
where $\psi_{s1}$ ($\psi_{s2}$) is the value of $\psi$ on the core surface, 
$r_1=a$ ($r_2=a$). This correctly reduces to (\ref{A12}) for 
$\ln 2\delta = {\tfrac{1}{2}}$. 

Consider now the flow as seen in the co-moving frame. This consists of two parts: 
an interior region composed of (non-circular) streamlines surrounding the vortices 
and an exterior region where the steamlines start and finish at $\infty$, where 
$\bb v = -U \bb 1_x$. The two regions are separated by an oblate dividing 
streamline, $x=\pm [{\rm tanh}(y/h)-y^2-{\tfrac{1}{4}}h^2]^{1/2}$, on which 
$\widetilde \psi = 0$; see the Figure on p.\ 221 of Lamb (1945). Here 
\begin{equation}
\widetilde \psi = -Uy + \psi = -{\frac{\kappa}{2 \pi}}\biggl({\frac{y}{h}}+\ln{\frac{r_1}{r_2}}\biggr)
\label{A19}
\end{equation}
is the streamfunction in the co-moving frame.

Since the interior fluid is, as seen in the laboratory frame, perpetually 
carried along by the vortex in its motion, it has a special significance. 
Its energy is [cf. (\ref{A18})]
\begin{equation}
\rho \kappa \widetilde \psi_{s1} = {\frac{\rho \kappa^2}{2 \pi}}\biggl(\ln{\frac{h}{a}}-{\frac{1}{2}}\biggr) ={\frac{\rho \kappa^2}{2 \pi}} \ln{\frac{h}{2a\delta}}=\widetilde {\cal E}.
\label{A20}
\end{equation}
The same argument applies with minor modifications to vortices with other 
internal structure.

\vfill\eject
\section*{References}

\noindent
 Berloff N G and Roberts P H 2000 {\it J Phys A: Math Gen} {\bf 33} 4025--38 

\noindent
 Crow S C 1970 {\it AIAA} {\bf8} 2172--9

\noindent
 Donnelly R J 1991 {\it Quantized Vortices in Helium II} (Cambridge:
Cambridge University Press)

\noindent
 Ford R and Llewellyn Smith S G 1999 {\it J Fluid Mech} {\bf 386} 305--28

\noindent
 Jones C A and Roberts P H 1982 {\it J Phys A: Gen Phys} {\bf 15} 2599--618

\noindent
 Jones C A, Putterman S J and Roberts P H 1986 {\it J Phys A: Math Gen} {\bf 19} 2991--3011

\noindent
 Koplik J and Levine H 1993 {\it Phys Rev Lett} {\bf 71} 1375--79

\noindent
 Koplik J and Levine H 1996 {\it Phys Rev Lett} {\bf 76} 4745--48

\noindent
Kuznetsov E A and  Juul Rasmussen J  1995 {\it Phys. Rev. E} {\bf 51} 4479--4484

\noindent
Lamb H 1945 {\it Hydrodynamics}, 6th edition (Dover Publications, New York)

\noindent
 Leadbeater M, Winiecki T, Samuels D C, Barenghi C F, 
and Adams C S 2001 {\it Phys Rev Lett} {\bf 86} 1410--13

\noindent
 Moore D W 1972 {\it Aero Quart} {\bf 23} 307--14

\noindent
Pitaevskii L P 1961 {\it Sov Phys JETP} {\bf 13} 451--54

\noindent
 Raymond G W and Kuo H L 1984 {\it Q J R Meteorol Soc} {\bf 110} 525--51

\noindent
Roberts P H and Donnelly R J 1970 {\it Phys Lett} {\bf 31A} 137--8

\noindent
Roberts P H 1972 {\it Mathematica} {\bf 19} 169--79

\noindent
 Saffman P G 1992 {\it Vortex Dynamics} (Cambridge: 
Cambridge University Press)

\noindent
 Schwarz K W 1988  {\it Phys Rev B} {\bf 38} 2398--417

\noindent
 Vinen W F 2000 {\it Phys Rev B} {\bf 61} 1410--20

\end{document}